\documentclass[prb,amssymb,twocolumn]{revtex4}
\usepackage{graphicx}

\begin{document}

\title{Field-induced Gap and Quantized Charge Pumping in Nano-helix}
\author{Xiao-Liang Qi$^{1,2}$ and Shou-Cheng Zhang$^{1}$}

\affiliation{$^1$Department of Physics, McCullough Building,
Stanford University, Stanford, CA 94305-4045}

\affiliation{$^2$Center for Advanced Study, Tsinghua University,
Beijing, 100084, China}
\begin{abstract}
We propose several novel physical phenomena based on nano-scale
helical wires. Applying a static electric field transverse to the
helical wire induces a metal to insulator transition, with the band
gap determined by the applied voltage. Similar idea can be applied
to ``geometrically" constructing one-dimensional systems with
arbitrary external potential. With a quadrupolar electrode
configuration, the electric field could rotate in the transverse
plane, leading to a quantized dc charge current proportional to the
frequency of the rotation. Such a device could be used as a new
standard for the high precession measurement of the electric
current. The inverse effect implies that passing an electric current
through a helical wire in the presence of a transverse static
electric field can lead to a mechanical rotation of the helix. This
effect can be used to construct nano-scale electro-mechanical
motors. Finally, our methodology also enables new ways of
controlling and measuring the electronic properties of helical
biological molecules such as the DNA.
\end{abstract}

\maketitle

Helical nanostructures occur ubiquitously in self-assembled form in
both inorganic
materials\cite{zhang2002,mcilroy2001,zhang2003,yang2004} and in the
biological world\cite{xu2004,endres2006}. In this paper, we propose
several novel physical phenomena based on nano-scale helix wires.
Firstly, when a uniform electric field is applied perpendicular to
the helical direction, the electrons moving in the nanowire
experience a periodical potential due to the potential energy
difference. Consequently, such a uniform electric field can induce a
gap in the electron energy spectrum, which drives the nanowire from
a metallic state to an insulating state if the electron density is
commensurate. The general principle behind such a simple phenomenon
is that a uniform electric field can generate a nonuniform potential
acting on a quantum wire if the quantum wire itself has a curved
shape. More generally, one can obtain a quasi one-dimensional system
in an arbitrary electric potential by applying an uniform electric
field to a quantum wire with proper shape. Recent advances in
nano-technology enables this design principle.

A more interesting phenomenon occurs when the applied electric field
is slowly rotated. When the system has a commensurate filling and
stays in the insulator state, a slow enough rotation of the electric
field satisfies the adiabatic condition and the system will stay in
the instantaneous ground state. During each period $T$ of the
electric field rotation, integer number of charge will be pumped
through the nanohelix, thus generating a quantized charge current.
In such a way, the nanohelix in a rotating electric field provides a
new realization of the quantum charge pumping effect proposed by
Thouless\cite{thouless1983}. The principle behind this charge
pumping effect is exactly the quantum analog of the celebrated
Archimedean screw invented more than two millenniums ago. By making
use of such an effect, one can design a quantized and controllable
current source. The reverse effect can also be studied, leading to
the possibility of a quantum nano-motor driven by electric current.

%Nano science is driven by several important technological goals.
%First of all, because of the exponential increase in the power
%density of the conventional CMOS chips, it is now widely recognized
%that new device concepts based on quantum principles and
%nano-structures are needed in order to extend the Moore's law.
%Secondly, there is increasing demand for multi-functionality on the
%same chip, integrating logic processing with electro-magnetic,
%electro-optic and electro-mechanical capabilities. Lastly, it is
%also desirable to control the basic properties such as the band gap,
%so that the color of light-emitting diodes and lasers can be
%continuously tuned. In this paper, we propose several device
%concepts based on nano-scale helix wires, which occurs ubiquitously
%in self-assembled form in both inorganic
%materials\cite{zhang2002,mcilroy2001,zhang2003,yang2004} and in the
%biological world\cite{xu2004,endres2006}.  This type of device has
%the potential of operating as transistor switches with lower power
%consumption and integrating multi-functionality within a single
%device. Inspirations for these device concepts are derived from the
%celebrated Archimedean screw invented more than two millenniums ago.

\begin{figure}[tbp]
\begin{center}
\includegraphics[width=3in] {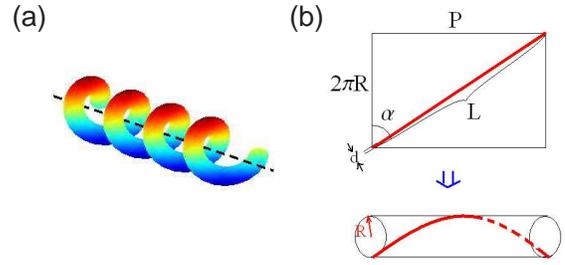}
\end{center}
\caption{(a) Schematic picture of a nanohelix. (b) The definition of
length scales $d,R,P,L$ shown for one pitch period of a helical
wire. \label{schematic}}
\end{figure}

\begin{figure}[tbp]
\begin{center}
\includegraphics[width=3in] {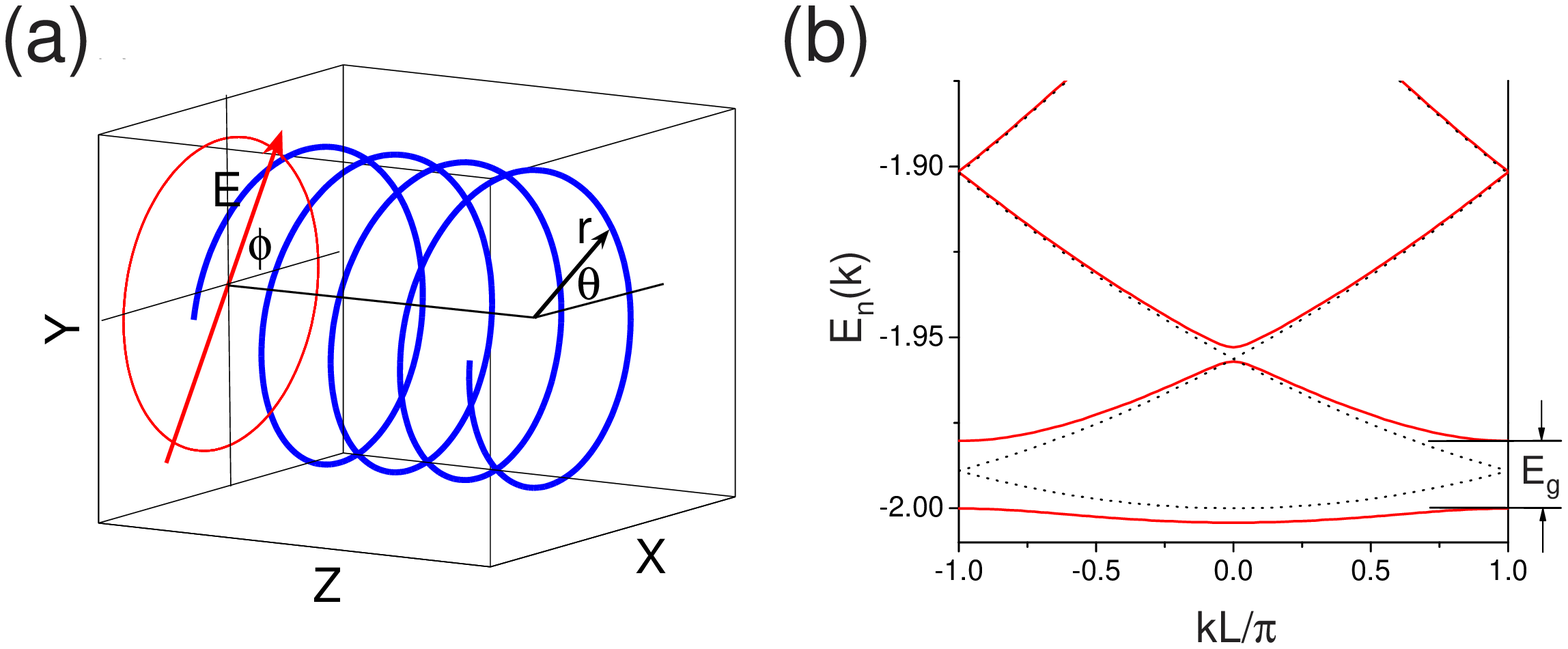}
\end{center}
\caption{(a) Definition of coordinates and direction of transverse
electric field. (b) Lowest 4 subbands under transverse field
$eER=0.02\hbar^2/2ma^2$ and $L=30a$, with $a$ being the lattice
constant. The dashed line shows the band structure in the reduced
zone scheme when $E=0$.} \label{chiralwire}
\end{figure}

To begin with, we consider a helical wire (as shown in Fig.
\ref{schematic}) where the diameter of the wire is $d$, the radius
of the helix is $R$, the helical angle is $\alpha$, the pitch length
of the helix is $P=2\pi R\tan\alpha$ and the net length of one
period is $L=2\pi R/\cos\alpha$. The carrier density of the wire is
denoted by $n$. In the present work, we will focus on the case when
the helical wire is sufficiently thin so that the electron motion
can be considered as one dimensional, and then discuss later in
detail the conditions imposed on the one-dimensionality.

In the continuum limit, the one-dimensional Hamiltonian of a
single electron in the nano-helix is simply written as %%
\begin{eqnarray} H_0=-\frac{\hbar^2}{2m}\partial_\rho^2\label{H0}\end{eqnarray}%%
with $m$ the effective mass and $\rho$ the length coordinate of the
helix. When a transverse electric field is applied, a potential
energy term is induced in the Hamiltonian. To write it down
explicitly, one can define the cylindrical coordinate system
$(r,\theta,z)$ as shown in Fig. \ref{chiralwire} (a), with the $z$
axis defined as the axis of the helix. The coordinate of a point
with length coordinate $\rho$ is \begin{eqnarray} r=R,\text{
}\theta=\frac{2\pi \rho}{L}({\rm mod}2\pi),\text{
}z=\rho\sin\alpha.\end{eqnarray}

\begin{figure}[tbp]
\begin{center}
\includegraphics[width=2.8in] {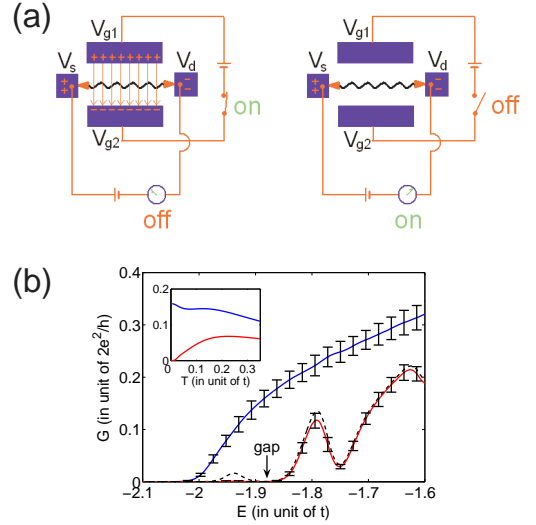}
\end{center}
\caption{(a) Illustration of the on and off state of the nano-helix
switch. The gate voltage is given by $V_g=V_{g1}-V_{g2}$ and
source-drain voltage is given by $V_{sd}=V_s-V_d$. (b) DC
conductance of the nano-helix connected with two
metallic leads. The blue and red line stands for the case with
$eER=0$ and $eER=0.2t$, respectively, where the error bar stands for
the impurity effect with the impurity potential $W=0.05t$. The
dashed line is the conductance without impurity under the same
electric field $eER=0.2t$. The arrow marks the first gap induced by
electric field. All the calculations are done under the temperature
$T=0.01t$ for the helix system with $N=200,~L=20,~\Gamma/2\pi=0.1t$.
The inset shows the temperature dependence of conductance with the
electric field $eER=0$ (blue line) or $eER=0.2t$ (red line).}
\label{switch}
\end{figure}

{\em ---The Quantum Helical Transistor (QHT).} When a transverse
uniform electric field ${\bf E}=E\left(\cos\phi\hat{\bf
x}+\sin\phi\hat{\bf y}\right)$ is applied, as shown in Fig.
\ref{chiralwire} (a), the potential energy is given by
\begin{eqnarray} H_{\rm el}=e{\bf E\cdot
r}(\rho)=eER\cos\left(\frac{2\pi
\rho}L-\phi\right).\label{Hamiltonian}\end{eqnarray} Thus the
combined single electron Hamiltonian takes the form of $H=H_0+H_{\rm
el}$. In this case, the continuous quadratic energy spectrum will
split into energy bands, with first Brillouin zone
$k\in[-\pi/L,\pi/L)$. The typical band structure is shown in Fig.
\ref{chiralwire} (b). The gap between the $n$th and $(n+1)$th bands
can be calculated by perturbation theory as $E_{g}(n)\propto
eER\left(eER/E_0\right)^{n-1}$ in the limit $eER\ll E_0\equiv
\hbar^2\pi^2/2mL^2$. For concreteness, we shall focus on the first
gap $n=1$, since it is the largest, corresponding to $E_g\simeq
eER$. The transverse electric field can be generated by applying a
voltage $V_g=V_{g1}-V_{g2}$ on the gate electrodes as shown in Fig.
\ref{switch} (a). Assuming that the helical wire just fits into the
gate electrodes, the resulting electric field is $E=V_g/2R$, and
therefore the band gap is simply given by $E_g\simeq eV_g/2$, which
is independent of the radius $R$ of the helix. On the other hand,
the average potential $V_a=(V_{g1}+V_{g2})/2$ relative to the
source-drain potential $(V_s+V_d)/2$ can be used to tune the
chemical potential of the wire and thus the electron density $n$.
When the chemical potential lies in the first gap, the system is an
insulator and the corresponding one-dimensional filling fraction is
$n_{\rm 1d}=2/L$, that is, two electrons per helical period. The
factor of two arises from the spin degeneracy.

%\begin{figure}[tbp]
%\begin{center}
%\includegraphics[width=2.5in] {band1d.eps}
%\end{center}
%\caption{Lowest 4 subbands with $eER=0.02\hbar^2/2ma^2$ and
%$L=30a$, with $a$ being the lattice constant. The dashed line
%shows the band structure in the reduced zone scheme when $E=0$.}
%\label{band1d}
%\end{figure}

Since the system with such a filling is gapless and conducting
before applying electric field, the transverse electric field leads
to a metal-insulator transition in the nano-helix, and thus defines
a new type of nano-scale transistor switch, the status of which is
``on" when the electric field is turned off, and ``off" when
electric field is turned on, as illustrated in Fig. \ref{switch}
(a). Such a switch can work under a source-drain voltage
$V_{sd}<V_g/2$, so that the chemical potential of both leads lie
inside the gap.

To make our discussions here and below more concretive, we can
regularize the Hamiltonian (\ref{H0}) to a tight-binding model:
\begin{eqnarray}
H_{\rm
TB}&=&-t\sum_{i=1}^{N-1}\left(c_i^\dagger c_{i+1}+h.c.\right)\label{Htb}\\
& &+eER\sum_{i=1}^N\cos\left(\frac{2\pi
a}{L}i-\phi\right)c_i^\dagger c_i+\sum_{i=1}^N\epsilon_ic_i^\dagger
c_i\nonumber
\end{eqnarray}
in which $a$ is the lattice constant and $N$ is the total number of
sites. The last term stands for quenched disorder, with the random
potential
$\left\langle{\left\langle{\epsilon_i\epsilon_j}\right\rangle}\right\rangle=\delta_{ij}W^2$.
To study the transport properties in such a mesoscopic system, we
need to include the effect of leads, which in this one-dimensional
model can be described by the terms below:
\begin{eqnarray}
H_{\rm Lead}&=&\frac{V}{\sqrt{\Omega}}\sum_{\bf
k}\left(c_{1}^\dagger a_{L{\bf k}}+c_{N}^\dagger
a_{R{\bf k}}+h.c.\right)\nonumber\\
& &+\sum_{{\bf k}}\sum_{\alpha=L,R}\left(\epsilon({\bf
k})-\mu_\alpha\right)a_{\alpha {\bf k}}^\dagger a_{\alpha {\bf
k}}\label{Hlead}
\end{eqnarray}
with $a_{L{\bf k}}$ and $a_{R{\bf k}}$ stands for the annihilation
operators of electron in the left and right leads, respectively.
Then the conductance can be calculated
by\cite{landauer1957,meir1992}
\begin{eqnarray}
G(E)&=&\frac{2e^2}{h}\int dE\left(-\frac{\partial f(E)}{\partial E}\right)\left|{t_{1,-1}(E)}\right|^2\\
t_{\alpha\beta}(E)&=&\Gamma\left\langle i_\beta\right|G^r(E)\left|
i_\alpha\right\rangle
\end{eqnarray}
in which $\Gamma=2\pi V^2n_F$ with $n_F$ the density of state on the
fermi surface of each lead, and a factor $2$ from spin degeneracy
has been included. For later convenience, the scattering amplitude
$t_{\alpha\beta}$ is defined, in which
$\left|i_1\right\rangle=\left|1\right\rangle,~\left|i_{-1}\right\rangle=\left|N\right\rangle$
are the local Weinner states on the left and right end site of the
nanowire, respectively. $G^r(E)$ is the retarded Green function of
the nanohelix, $G^r(E)=\left[E+i\delta-H_{\rm
TB}-\Sigma\right]^{-1}$. Under wide band approximation, the
self-energy is $\Sigma=-\frac
i2\Gamma\left(\left|1\right\rangle\left\langle
1\right|+\left|N\right\rangle\left\langle N\right|\right)$.

%\begin{figure}[tbp]
%\begin{center}
%\includegraphics[width=3in] {Gdc.eps}
%\end{center}
%\caption{DC conductance of the nano-helix connected with two
%metallic leads. The blue and red line stands for the case with
%$eER=0$ and $eER=0.2t$, respectively, where the error bar stands for
%the impurity effect with the impurity potential $W=0.05t$. The
%dashed line is the conductance without impurity under the same
%electric field $eER=0.2t$. The arrow marks the first gap induced by
%electric field. All the calculations are done under the temperature
%$T=0.01t$ for the helix system with $N=200,~L=20,~\Gamma/2\pi=0.1t$.
%The inset shows the temperature dependence of conductance with the
%electric field $eER=0$ (blue line) or $eER=0.2t$ (red line).}
%\label{Gdc}
%\end{figure}

The typical behavior of conductance is shown in Fig. \ref{switch}
(b), in which the conductance with and without external electric
field is compared. The metal-insulator transition induced by the
electric field can be seen explicitly from the temperature
dependence of the conductance, as shown in the inset of fig.
\ref{switch} (b). Another important information from this
calculation is that weak impurity $W\ll eER$ can further widen the
insulating region induced by electric field, since the
electric-field induced subband is much narrower than the original
energy band in the helix, and thus much easier to be localized by
disorder. As shown in fig. \ref{switch} (b), under the same disorder
strength, the first subband of the system with electric field is
fully localized, while the one without electric field remains
metallic.

%\begin{figure}[tbp]
%\begin{center}
%\includegraphics[width=2.5in] {pumping.eps}
%\end{center}
%\caption{Illustration of the quantized charge pumping effect, with
%four electrodes causing a rotating electric field.} \label{pumping}
%\end{figure}

However, a strong disorder $W\gtrsim eER$ can dominate the effect of
electric field and thus kill this metal-insulator transition.
Another important issue in this system is the elecron-electron
interaction. According to the Luttinger liquid theory, repulsive
interaction will make periodical potential more relevant and thus
further stabilize this switch effect\cite{kane1992}. In summary, the
terms that may harm this effect are attractive interaction and
strong impurity. Although we won't involve more quantitative
discussion in the present paper, a lower-limit estimate to the
stability of the present effect can be given as $W\ll E_g$, $V\ll
E_g$, with $W, V$ the characteristic energy scale of impurity random
potential and phonon-induced attractive interaction, respectively.
Under such a condition, the electric field-induced potential
scattering dominant the interaction and impurity effect and thus the
switching effect (and also the charge pumping and motor effect shown
below) remains robust.

\begin{figure}[tbp]
\begin{center}
\includegraphics[width=2.5in] {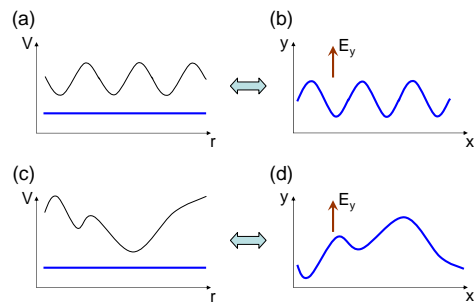}
\end{center}
\caption{Schematic illustration of the geometrical design of
one-dimensional potentials. (a) A straight quantum wire in a sine
wave potential (blue line) is equivalent to (b) a periodically
curved quantum wire in a uniform electric field ${\bf E}$. (c) More
generally, a straight quantum wire in an arbitrary potential $V(r)$
(blue line) is equivalent to (d) a curved quantum wire in a uniform
electric field ${\bf E}$. } \label{geodesign}
\end{figure}

The device concepts discussed so far depend only on the periodically
curved nature of the helical wire, and does not depend on the net
helicity of the wire. Therefore, these concepts can be equally well
implemented by patterning a quasi-1D wire in a periodically curved
form, e.g. a sine wave form, on a plane, and by applying a
transverse voltage. The helical wire perhaps has the advantage of
being self-assembled and can be more easily realized in the
nano-scale. In principle, the same idea can be generalized to design
an arbitrary potential in a quasi-one-dimensional system. Consider a
planar quantum wire with the shape of function $y(x)$ in an uniform
electric field ${\bf E}=E{\bf \hat{y}}$, then the effective
one-dimensional potential $V(r)$ is determined by
\begin{eqnarray}
\frac{Ey'(x)}{\sqrt{1+y'(x)^2}}=-\frac{dV}{dr}
\end{eqnarray}
in which $r$ is the arc length of the wire. In this way, one can
obtain a quasi-one-dimensional system in any potential $V(r)$ by
choosing a proper shape $y(x)$, as shown in Fig. \ref{geodesign}.
Such a "geometrical design" of one-dimensional systems takes the
advantage of tunable strength and shape of potential energy, and
thus can help to produce artificial one-dimensional materials with
highly controllable electronic and optical properties. In
particular, our device can realize a light-emitting-diode (LED) with
tunable bandgap and color, controlled purely by the external gate
voltage. %{\it draw in fig 4 first the sine wave design, then the
%arbitrary shape design.}

%\begin{figure*}[tbp]
%\begin{center}
%\includegraphics[width=2in] {helixshow1.eps}\includegraphics[width=2in] {helixshow2.eps}\includegraphics[width=2in] {helixshow3.eps}
%\end{center}
%\begin{center}
%\includegraphics[width=2in] {helixshow4.eps}\includegraphics[width=2in] {helixshow5.eps}
%\end{center}
%\caption{Nanohelix with integer filling factor in an external
%electric field. The darker color represents higher electron
%density. Following a full rotation of electric field from (a) to
%(e), the electron density distribution returns to itself, but with
%an integer number of electrons pumped from left to right over one
%helical pitch distance. } \label{helixshow}
%\end{figure*}

{\em --The Quantum Helical Pump (QHP).} We now consider an
adiabatical rotation of the transverse electric field, when a more
interesting effect emerges in the nano-helix system. Experimentally,
the rotation of electric field can be realized by a set of
quadrupolar electrodes, as shown in Fig. \ref{Gpump} (a). A rotating
electric field with angular frequency $\omega$ is described by a
time-dependent $\phi(t)=\omega t$ in the Hamiltonian
(\ref{Hamiltonian}). In the adiabatical limit $\hbar\omega\ll E_g$,
the gapped system with commensurate filling $n_{\rm 1d}=2/L$ will
stay in the time-dependent ground state. Similar to what
Thouless\cite{thouless1983} proposed by using a sliding linear
periodic potential, such an adiabatical translation of periodical
potential on a gapped electron system can in general lead to a
quantized charge pumping current
\begin{eqnarray}
J=2Ne\frac{\omega}{2\pi},
N\in\mathbb{Z},\label{current}
\end{eqnarray}
in the zero temperature limit, which means $N$ electrons per spin
component are pumped through the wire system during each period
$T=2\pi/\omega$. %In the nano-helix system with one subband filled
%(or $n_{\rm 1d}=2/L$), direct calculation shows that $N=1$.
Intuitively, such a quantized charge pumping can be easily
understood as a quantum version of Archimedean screw. Due to the
electric force, the electron density in the lower subband is larger
on the side nearer to positive electrode, and a charge-density-wave
(CDW) is induced by the transverse electric field. Consequently, the
high-density region will follow the rotation of electric field and
thus the coordinate of each electron shifts by one pitch distance
during one period of electric field.

More quantitatively, the current $J_{\rm pump}$ induced by a
time-dependent electric field can be calculated in the tight-binding
model (\ref{Htb}) and (\ref{Hlead}) in a similarly way as the DC
conductance:\cite{brouwer1998,aharony2002}
\begin{eqnarray}
G_{\rm pump}&\equiv&\frac{J_{\rm
pump}}{e\omega}\nonumber\\
&=&2\int
\frac{dE}{2\pi}\int_0^{2\pi}\frac{d\phi}{2\pi}\left(-\frac{\partial
f(E)}{\partial
E}\right)\nonumber\\
& &\cdot\sum_{\alpha,\beta=\pm 1}\beta{\rm
Im}\left[t_{\alpha\beta}^*(E,\phi)\frac{\partial
t_{\alpha\beta}(E,\phi)}{\partial \phi}\right]\label{Gpump}
\end{eqnarray}

A typical result of this calculation is shown in Fig. \ref{Gpump}
(b). As expected by topological protection, random disorder can only
induce fluctuation of $G$ for gapless system, and leaves the
quantized plateaus unchanged. Actually, under zero temperature such
a quantized adiabatical charge pumping is robust under any
deformation of the Hamiltonian, as long as the subband gap $E_g$ is
not closed. In particular, even if the two AC voltages applied to
the quadrupolar electrodes are not perfectly harmonic but with some
deformations or noises, as long as the electric field vector ${\bf
E}(t)$ still encircles the $(0,0)$ point once each period, the
quantization of pumping conductance (in the zero temperature limit)
remains robust {\em without any correction}. In the same way it will
remain robust when the nano-helix has a different shape as shown in
Fig.\ref{schematic} (a) but with the same helical topology.

The pumping conductance $G_{\rm pump}$ at finite temperature is
simply a convolution of the zero temperature result $G_{\rm
pump}(T=0)$ with the thermo factor $-\partial f(E)/\partial E$.
Consequently, $G_{\rm pump}$ will deviate from the quantized value.
However, for a quantized plateu $G_{\rm pump}(T=0)=2N/2\pi$ with
width $E_g$, the deviation $\delta G=G_{\rm pump}(T)-G_{\rm
pump}(T=0)$ at the middle-point of the plateu can be estimated by
$\delta G/G_{\rm pump}(T=0)\simeq -\frac{2}{e^{\beta E_g/2}+1}$,
which is exponentially small when $k_BT\ll E_g$.

Compared with the earlier experiments to realize Thouless's charge
pumping effect, like those involving surface accoustic
wave\cite{shilton1996,aharony2002} or deformation potential on a
quantum dot\cite{switkes1999}, the present realization has the
advantage of ``coding" the topological information directly into
the geometrical structure of the self-assembled nano-helix, whose
long periodic structure makes the effect more intrinsic and
robust. Our device could have higher precision of the current
quantization and potentially lead to a new standard of current
definition.\cite{niu1990}
\begin{figure}[tbp]
\begin{center}
\includegraphics[width=2.5in] {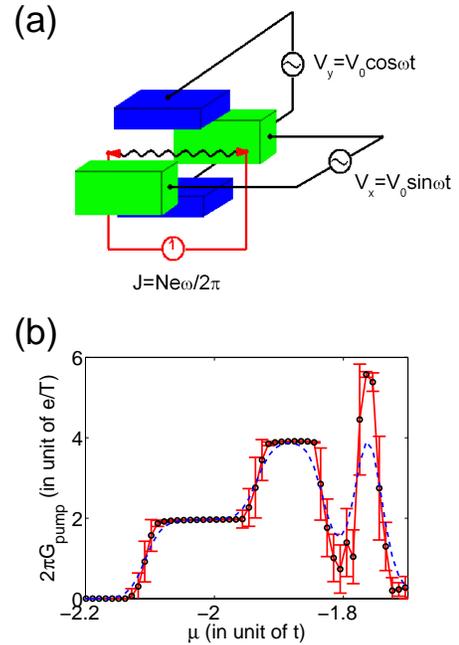}
\end{center}
\caption{(a) Illustration of the quantized charge pumping effect,
with four electrodes causing a rotating electric field. (b) Pumping
conductance $G_{\rm pump}=J_{\rm pump}/e\omega$ under zero
temperature(red solid line) and finite temperature $T=0.01t$ (blue
dashed line). The error bar shows the fluctuation induced by the
disorder strength $W=0.1t$. The parameters of the tight-binding
system are taken as $N=100,~L=20,~eER=0.2t,\Gamma/2\pi=0.1t$.}
\label{Gpump}
\end{figure}

\begin{figure}[tbp]
\begin{center}
\includegraphics[width=1.5in] {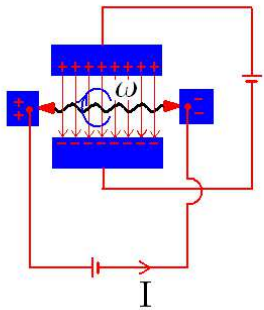}
\end{center}
\caption{Illustration of the quantum helical nano-motor. }
\label{motor}
\end{figure}

{\em ---The Quantum Helical Motor (QHM)} As a direct inverse effect
of the topological charge pumping, a nano-helix in a transverse
electric field can work as a quantum motor, where a longitudinal
current can lead to an uniform mechanical rotation with the
frequency, as shown in Fig. \ref{motor}
\begin{eqnarray}
\omega=\frac{\pi J}{Ne},N\in
\mathbb{Z}.\label{frequency}\end{eqnarray} This is a direct quantum
analog of a propeller or a windmill. In order to realize this
effect, both ends of the helical wire should be attached to some
kind of molecular swivel, similar to those described in
Ref.\cite{bryant2003}, which enables the uni-axial rotation of the
helix. It is also possible to drive an AC current $J(t)$ through a
helical wire with fixed ends, which will cause an AC oscillation of
the helix. However, the AC effect is not as robust as DC effect,
since the relation between AC oscillation and AC current is
generally not protected by topological reason.
%On the other hand, if
%both ends of the helix are fixed, it is still possible to drive an
%AC current $J(t)$ through the helical wire, such that an AC
%oscillation of the helix is induced, whose angular velocity is given
%by $\omega(t)=\pi J(t)/Ne$.

The relation (\ref{frequency}) is generally true under any friction
or other perturbations, as long as $\hbar\omega\ll E_g$ and $k_BT\ll
E_g$ so that the adiabatical evolution condition is satisfied. When
there are more frictions, it will be harder to inject a current, but
the relation between frequency and current remains valid. In the
extreme case, if the nano-helix is fixed, then $\omega=0$ and at the
same time $J=0$, which recovers the switch effect. Suppose there is
a frictional torque $\mathcal{T}=-\eta \omega$ acting on the helix,
then the energy cost per unit time is
$P=-\mathcal{T}\omega=\eta\omega^2$. Consequently, one needs a
finite voltage $V$ to drive a constant current in this helix. The
voltage is determined by the energy equilibrium condition
$P=\eta\omega^2=VJ$, which implies that the power of the voltage
cancels the friction energy cost. Thus we get the relation
\begin{eqnarray} \eta\left(\frac{\pi J}{Ne}\right)^2=VJ\Rightarrow
R=\frac{V}{J}=\frac{\pi^2\eta}{N^2e^2},\label{resistance}\end{eqnarray}
which relates friction to a resistivity. As has been discussed in
switch effect, the source-drain voltage $V$ must satisfy $V<E_g/e$
so as to keep the adiabatical evolution. Consequently, for a given
friction $\eta$, the rotating frequency of such a nano-motor is
restricted by $\omega<\frac{\pi}{Ne}\frac{V_{\rm
max}}R=\frac{E_gN}{\pi\eta}$ and also by the adiabatical condition
$\hbar\omega\ll E_g$.

{\em ---More discussions on experimental realizations}

After proposing these three effects, we now analyze the detailed
experimental conditions for their realizations:

\begin{enumerate}
\item{The system is quasi one-dimensional, which requires $E_g\ll
E_\perp$, $E_F\ll E_\perp$, with $E_\perp$ the transverse excitation
gap and $E_F=k_F^2/2m$ the fermi energy. }\item{The electron (or
hole) filling is commensurate $n_{\rm 1d}=2N/L, N\in \mathbb{N}$. }
\item{The adiabatical approximation is applicable, which requires (i) temperature $k_BT\ll E_g$; (ii) impurity and
attractive interaction energy scale $W,U\ll E_g$; (iii) rotation
frequency of electric field or nano-helix $\omega\ll E_g/\hbar$. }
\item{The total length of the nanohelix $L_{\rm tot}\gg\xi= \frac{\hbar
v_F}{E_g}$, so as to prevent the direct tunneling between the two
ends and protect the topological transport. }
\end{enumerate}

To satisfy the requirements above, an ideal nano-helix for our
purpose should have thin diameter $d$, large helix radius $R$, long
length $L_{\rm tot}$ and also be very clean. Experimentally, two
most possible candidates for this effect are helical nano-wires made
from ZnO, SiC, CB,
etc.\cite{zhang2002,mcilroy2001,zhang2003,yang2004} and chiral
biological molecules such as RNA, DNA and some proteins. To make the
discussion more concrete, here we give an estimate of the present
effects in the deformation-free ZnO nano-helix realized in
Ref.\cite{yang2004}. The size of that nano-helix is reported as
$d\simeq 12{\rm nm}$, $\alpha\simeq 40^\circ$, $R=15{\rm nm}$,
$L=6R/\cos\alpha\simeq 123{\rm nm}$ (the estimate of $L$ is a little
different from the previous one since the intercept of ZnO helix
here is hexagonal rather than round.) If we approximate the electron
effective mass by the bulk ZnO value $m\simeq
0.24m_e$\cite{karpina2004}, then the transverse excitation gap can
be estimated as $E_\perp\sim \frac{h^2}{2m d^2}\simeq 44{\rm meV}$.
The filling corresponding to first gap is $n_{\rm 1d}=2/L\sim
1.6\times 10^5/{\rm cm}$, which in 3-d unit gives $n_{\rm
3d}=2/L\pi(d/2)^2\simeq 1.4\times 10^{17}/{\rm cm}^3$. The
corresponding $E_F=\hbar^2 k_F^2/2m\simeq 0.1{\rm meV}$. Thus the
condition $E_F\ll E_\perp$ is always satisfied, and condition
$E_g\ll E_\perp\simeq 44{\rm meV}$ requires the electric field
$E\ll3\times 10^6{\rm V/m}$ or gate voltage $V_g\ll 88{\rm mV}$. The
$\xi$ in condition (4) is $\xi=\hbar^2 k_F/mE_g\simeq 1.6{\rm nm}$,
thus the condition (4) is always satisfied. If we take an electric
field $E=3.3\times 10^5{\rm V/m}$ or $E_g=eER=5{\rm meV}$, then the
conditions 3 requires i) $T\ll 60K$; ii) in condition 3, $W,U\ll
5{\rm meV}$; iii) $\omega\ll 7.6\times 10^{12}{\rm Hz}$. In summary,
such an effect should be observable in a wide temperature range for
the ZnO nanowire in Ref.\cite{yang2004}, if it is clean enough and
the doping can be controlled well. (To avoid impurity effect, the
filling should be controlled by gate voltage rather than chemical
doping. )

Compared to the inorganic nano-helixes, chiral biological
molecules such as RNA, DNA or protein may have the advantage of
better one-dimensionality, which implies a larger transverse gap
$E_\perp$, since they can be much thinner than the nano-wires. For
the effects proposed here to be observed, one needs to find
molecules which are semi-conducting and have a good
one-dimensional energy band. In a recent review
article\cite{endres2006}, transport properties of various DNA
molecules are summarized, some of which show semiconducting
behavior. If for some molecules $E_\perp\gtrsim 1{\rm eV}$, then
it's possible to observe the effect at room temperature with
voltage $V_g\gtrsim 200{\rm meV}\gg 2k_BT\simeq 60{\rm meV}$.

In summary, in this paper we proposed three related effects in
quantum helical systems under a transverse electric field. Under a
slowly rotating electric field, a nanohelix with commensurate
filling works as a quantum Archimedean screw. The experimental
conditions to realize such effects are shown to be feasible for
present experimental techniques. Since helical structures occur
naturally in the biological world, the principles discussed here
also provides new methods to control and detect biological
molecules.
%could also enable the construction of
%large scale, integrated, multi-functional biological systems for
%information processing.

{\bf Acknowledgement.} The authors wish to thank B. A. Bernevig, D.
Cox, Y. Cui, S. Doniach, C. Huang, T. Hughes, S. Kivelson, C.-X.
Liu, P. Wong and B.-H. Yan for useful discussions. This work is
supported by the NSF under the grant No. DMR-0342832 and the US
Department of Energy, Office of Basic Energy Sciences under contract
No. DE-AC03-76SF00515.

\bibliography{chargepumping}
%\begin{thebibliography}{10}
%
%\bibitem{zhang2002}
%H.-F. Zhang, C.-M. Wang, and L.-S. Wang, Nano Lett. {\bf 2},  941
%(2002).
%
%\bibitem{mcilroy2001}
%D.~N. McIlroy, D. Zhang, Y. Kranov, and M.~G. Norton, Appl. Phys.
%Lett. {\bf
%  79},  1540  (2001).
%
%\bibitem{zhang2003}
%D. Zhang {\it et~al.}, Nano Lett. {\bf 3},  983  (2003).
%
%\bibitem{yang2004}
%R. Yang, Y. Ding, and Z.~L. Wang, Nano Lett. {\bf 4},  1309  (2004).
%
%\bibitem{xu2004}
%B. Xu, P. Zhang, X. Li, and N. Tao, Nano Lett. {\bf 4},  1105
%(2004).
%
%\bibitem{endres2006}
%R.~G. Endres, D.~L. Cox, and R.~R.~P. Singh, Rev. Mod. Phys. {\bf
%76},  195
%  (2006).
%
%\bibitem{landauer1957}
%R. Landauer, IBM J. Res. Dev. {\bf 1},  233  (1957).
%
%\bibitem{kane1992}
%C.~L. Kane and M.~P.~A. Fisher, Phys. Rev. B {\bf 46},  15233
%(1992).
%
%\bibitem{thouless1983}
%D.~J. Thouless, Phys. Rev. B {\bf 27},  6083  (1983).
%
%\bibitem{thouless1982}
%D.~J. Thouless, M. Kohmoto, M.~P. Nightingale, and M. den Nijs,
%Phys. Rev.
%  Lett. {\bf 49},  405  (1982).
%
%\bibitem{shilton1996}
%J.~M. Shilton {\it et~al.}, J. Phys.: Condens. Matter {\bf 8},  L531
%(1996).
%
%\bibitem{aharony2002}
%A. Aharony and O. Entin-Wohlman, Phys. Rev. B {\bf 65},  241401
%(2002).
%
%\bibitem{switkes1999}
%M. Switkes, C.~M. Marcus, K. Campman, and A.~C. Gossard, Science
%{\bf 283},
%  1905  (1999).
%
%\bibitem{niu1990}
%Q. Niu, Phys. Rev. Lett. {\bf 64},  1812  (1990).
%
%\bibitem{bryant2003}
%Z. Bryant {\it et~al.}, Nature {\bf 424},  338  (2003).
%
%\bibitem{karpina2004}
%V.~A. Karpina {\it et~al.}, Crystal Research and Technology {\bf
%39},  980
%  (2004).
%
%\end{thebibliography}

\end{document}